\documentclass[prx,twocolumn,groupedaddress,floatfix]{revtex4-2}

\usepackage{amsmath}
\usepackage{amssymb}
\usepackage{graphicx}
\graphicspath{{./}}
\usepackage{hyperref}
\usepackage{placeins}

\begin{document}

\title{Challenges in the simulation of enzymatic transition states
with emerging multireference character}

\author{Matthias Kaiser}
\author{Sanjoy Ray}
\author{Valentin Kasper}
\email{valentin@pexmachina.com}
\affiliation{PexMachina Inc., 8 The Green, STE R,
Dover, DE 19901, USA}

\date{\today}

\begin{abstract}
Highly successful drugs, such as transition-state inhibitors,
were designed by mimicking the precise geometry of enzymatic
transition states, a strategy that requires quantum chemistry
accurate enough to specify molecular structure at
sub-\AA\ resolution.
A central challenge during such a quantum chemistry simulation
is that classical single-reference method can deviate
substantially from the exact energy at strongly correlated
transition states.
We study the breakdown due to multireference character
by studying a model system with a multireference
transition state.
In this work we present a proprietary PexMachina solver which
reproduces the exact ground-state energy throughout the
transition state scan, while single-reference methods
struggle for certain geometries to achieve chemical accuracy.
\end{abstract}

\maketitle

\section{Introduction}
\label{sec:intro}

Enzyme-catalyzed reactions determine the pharmacokinetics of
many approved drugs~\cite{Silverman2004} and the biosynthesis
of natural products~\cite{Dewick2009},
and achieving the quantum-chemical accuracy needed to
reliably predict catalytic rates and guide drug
design is essential at this scale.
The activation barrier and transition-state (TS) geometry
are the decisive quantities for predicting reaction rates
and designing drugs: a 1~kcal/mol error in the
free-energy barrier leads to an order-of-magnitude error
in the predicted rate at room
temperature~\cite{Schramm2011,Wolfenden1969}, and a
bond-length error of 0.05~\AA\ at the breaking bond
propagates directly into the shape and binding affinity
of the drug which should be developed~\cite{Schramm2011}.
Enzyme catalysis arises from geometric and electrostatic
complementarity to the transition state rather than to the
substrate: the active site is pre-organized to stabilize
the TS structure~\cite{Pauling1948}.

Building on this principle, transition-state analogues,
which are molecules designed to mimic the TS geometry
and charge distribution, bind with picomolar affinity,
orders of magnitude tighter than any substrate
mimic~\cite{Schramm2011}.
When the enzyme active-site structure is available from
crystallography, a TS model can be constructed from
structural and mechanistic information without quantum
chemistry: the HIV-1 protease crystal structure provided
the active-site template from which the TS geometry was
inferred, guiding the design of saquinavir and related
antiretrovirals~\cite{Navia1989,Wlodawer1989}.
However, for many enzyme targets no transition-state
analogue with sufficient geometric fidelity has been
crystallized, and the TS geometry must instead be obtained
from quantum chemical computation.
Choosing the right quantum chemical method is, however,
non-trivial: methods differ by many orders of magnitude
in cost and by chemically significant amounts in
accuracy~\cite{Cramer2004}.
Obtaining a reliable TS geometry requires a method accurate
enough to resolve the barrier at chemical accuracy.
The relevant notion of accuracy is \emph{chemical accuracy},
corresponding to $1$~kcal/mol ($\approx 1.6\times10^{-3}$~Ha), the
threshold below which computed energy differences can usefully
guide experimental decisions~\cite{Cramer2004}.

Although systematic ab initio methods such as CCSD(T)
allow for precise modeling of transition states, they
become computationally prohibitive as the system grows:
canonical CCSD(T) scales as $\mathcal{O}(N^{7})$ with the
number of basis functions and is typically limited to
roughly 20\textendash30 heavy atoms in
practice~\cite{Raghavachari1989}.
Active-space methods such as CASSCF and CASPT2 face a
complementary bottleneck: their cost grows exponentially
with the number of correlated electrons in the active
space, restricting them to roughly 20 electrons in 20
orbitals without further
approximation~\cite{Roos1980,Andersson1992}.

This speed-accuracy trade-off shapes every methodological
choice: faster methods sacrifice rigor, while more accurate
methods incur steeply growing computational cost.
Among electronic structure methods, DFT sits at the
high-throughput end of this spectrum but may carry
systematic errors in the barrier
heights~\cite{Goerigk2011}.

When the system is large enough that CCSD(T) is
prohibitively expensive and the electronic structure is
sufficiently multireference that DFT and single-reference
coupled-cluster methods both break down, methods capable
of treating strong correlation exactly
become necessary~\cite{Roos1980,Andersson1992,White1992}.
However, exact methods such as FCI or CASSCF scale
exponentially with active-space size and are therefore
confined to small surrogate systems; they cannot be applied
directly to the large active spaces encountered in enzymatic
transition states.

In this study we focus on the multireference failure mode
by analyzing the H4 rectangular-to-square scan as a
minimal molecular surrogate for a transition state with
strong static correlation~\cite{Jankowski1980}.
The term \emph{strongly correlated} is used here in its
technical sense: the ground state cannot be described by
a single reference determinant, regardless of how that
determinant is chosen.
H4 is chosen precisely because its simplicity makes this
phenomenon analytically transparent at the four-electron
level; it demonstrates that the static-correlation
signature of a multireference transition state is already
present at this scale and serves as the first quantitative
check for any method claiming relevance to strongly
correlated enzymatic transition states.

This paper is organized as follows: in
Sec.~\ref{sec:tsi} we present transition-state inhibitors
as a concrete example of why accurate enzymatic TS computation
matters pharmaceutically;
Sec.~\ref{sec:model} introduces the model
system; Sec.~\ref{sec:methods} details the computational
methods; Sec.~\ref{sec:results} presents the results; and
Sec.~\ref{sec:discussion} discusses implications for
quantum simulation.

\section{Example: transition-state inhibitors}
\label{sec:tsi}

Enzymes achieve catalysis by being geometrically and
electrostatically complementary to the transition state, not
to the ground state; this principle~\cite{Pauling1948} implies
that TS analogues bind the active site with affinities orders
of magnitude tighter than any substrate mimic~\cite{Schramm2011}.
Transition-state inhibitors are drugs deliberately designed
to mimic this TS geometry and electronic structure, making
their design a direct pharmaceutical application of accurate
TS computation, and illustrating concretely why errors of
even 1~kcal/mol in the barrier translate into worse drug
candidates.

Enzymatic catalysis arises from TS complementarity~\cite{Pauling1948},
implying a quantitative relationship between catalytic power
and inhibitor affinity~\cite{Wolfenden1969,Lienhard1973,Schramm2011}:
\begin{equation}
  K_d \;\approx\; K_m \times
  \frac{k_\text{uncat}}{k_\text{cat}} ,
  \label{eq:tsa-affinity}
\end{equation}
where $K_d$ is the dissociation constant of the
transition-state analogue, $K_m$ is the Michaelis constant,
$k_\text{uncat}$ is the uncatalyzed rate constant in
solution, and $k_\text{cat}$ is the catalytic rate constant.
For efficient enzymes the rate enhancement is
$10^6$\textendash$10^{17}$, predicting TS analogue affinities in the
picomolar-to-femtomolar range, which are among the
tightest non-covalent binding affinities known in
biology~\cite{Schramm2011}; immucillin-H,
the purine nucleoside phosphorylase (PNP) inhibitor, binds
PNP with $K_d \sim 23$~pM compared with $K_m \sim 40~\mu$M
for the natural substrate, representing a
$\sim\!1700$-fold affinity gain over the best ground-state
analogue (formycin~A, $K_i \sim 40$~nM)~\cite{Schramm2011}.

Several further approved drugs were designed using the TS
complementarity principle, guided by crystallographic
active-site geometry rather than quantum-chemical TS
computation.
The HIV-1 protease crystal structure was resolved
independently by two groups~\cite{Navia1989,Wlodawer1989},
providing the geometric template for the hydroxyethylene
isostere core present in saquinavir~\cite{Vacca1994},
ritonavir~\cite{Kempf1995}, and lopinavir~\cite{Sham1998}.
Captopril was designed from knowledge of the active-site
zinc-coordination geometry and catalytic mechanism of
angiotensin-converting enzyme, producing a drug that
remains a cornerstone of hypertension
treatment~\cite{Ondetti1977}.
Zanamivir and oseltamivir were designed by mimicking the
oxocarbenium TS of sialic acid cleavage by influenza
neuraminidase~\cite{vonItzstein1993,Kim1997}.

One of the challenges in transition-state inhibitor design is
that the TS exists for only femtoseconds
($\sim\!10^{-14}$\textendash$10^{-13}$~s)~\cite{Schramm2011}
and cannot be observed directly; its structure must be
inferred from kinetic isotope effect (KIE)~\cite{Westheimer1961}
measurements and quantum chemical computation.
KIEs encode bond-order changes at the TS in isotope-dependent
rate ratios, while computation supplies the three-dimensional
geometry and electrostatic potential that KIEs alone cannot
fully determine.
These two approaches are mutually constraining: the
computational TS must reproduce measured KIEs before it can be
trusted as a design template.

Errors in the computed TS geometry propagate directly into
inhibitor shape and partial-charge complementarity, and
therefore into binding affinity.
Transition-state inhibitor design is thus a concrete,
commercially validated task where electronic structure accuracy
has direct, measurable pharmaceutical value.
The required accuracy---barrier heights to $\pm 1$~kcal/mol
and TS geometries to $\lesssim\!0.05$~\AA---must be achieved.

\begin{figure}[tbp]
\centering
\includegraphics[width=\columnwidth]{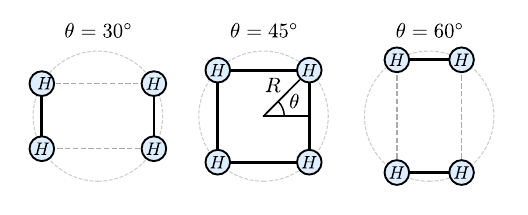}
\caption{%
  H$_4$ model of the H$_2$ + H$_2$ bond-exchange reaction.
  Four hydrogen atoms sit at the corners of a rectangle
  parameterized by angle~$\theta$, each at distance
  $R = 1.2$~\AA\ from the center.
  Solid bonds indicate the shorter, energetically favored
  H-H pairing at each geometry; dashed bonds indicate the
  longer, disfavored pairing.
  At $\theta < 45^\circ$ (left) the vertical pairs are closer;
  at $\theta > 45^\circ$ (right) the horizontal pairs are closer.
  At the square geometry $\theta = 45^\circ$ (center) both
  pairings are exactly degenerate: the ground state is a
  near-equal superposition of two configurations, making this
  the prototypical strongly correlated transition state
  for which single-reference methods break down.
}
\label{fig:h4-geometry}
\end{figure}

\section{Model}
\label{sec:model}

Enzyme active sites achieve their remarkable rate accelerations
through precisely arranged chemical environments, and
computational modeling has become indispensable for
understanding how they work~\cite{Senn2009,Siegbahn2009}.
The standard approach uses QM/MM or cluster models, in which
the reactive core is treated with a quantum-chemical method
while the surrounding protein is described at lower
cost~\cite{Cornell1995,Brooks1983}.
The reliability of any such study rests on the quality of
that QM method~\cite{Blunt2022}, which has motivated
the development of enzyme reaction benchmark
sets~\cite{Kromann2016,Sirirak2020,Wappett2019} and the
adoption of CCSD(T)~\cite{Raghavachari1989} as the reference
against which density functionals are judged.

The DFT and CCSD(T)-dominated QM/MM hierarchy is reliable for
enzymes whose chemistry is genuinely single-reference,
such as proton transfers and main-group organic
transformations, but it breaks down for
metalloenzymes where transition-metal centers introduce strong
static correlation that a single reference determinant cannot
capture~\cite{Goings2022}.
The Cu$_2$O$_2$ core of tyrosinase and hemocyanin provides
a well-studied example~\cite{Solomon2001,Cramer2006}.
Hemocyanin is the dinuclear copper protein that serves as
the oxygen carrier in mollusks and arthropods; its active
site binds O$_2$ as a side-on ($\eta^2:\!\eta^2$) peroxo
bridge across two Cu(II) ions, which can isomerize to a
bis($\mu$-oxo) Cu(III) form~\cite{Solomon2001}.
As the Cu$_2$O$_2$ core interconverts between these two
forms, the electronic structure passes through a strongly
biradical region where CCSD(T) exhibits poor convergence
and errors of tens of kcal/mol, while hybrid DFT and
single-root CASPT2 err in opposite
directions~\cite{Cramer2006,Flock1999,Rode2005},
and only multireference treatments recover the correct
profile~\cite{Kurashige2011,Kurashige2014}.

This breakdown is not specific to metal chemistry but arises
whenever the electronic structure approaches near-degeneracy.
In the forbidden [2+2] exchange of two H$_2$ molecules,
the four atoms pass through a square transition state where
the HOMO and LUMO become degenerate and the singlet ground
state acquires strong biradical character, with the leading
natural-orbital occupation depressed from 2 to approximately
1.74 and the lowest correspondingly raised to approximately
0.26~\cite{Jankowski1980}.
At that geometry, restricted DFT and RHF overestimate the
barrier, CCSD(T) undershoots it through its non-variational
coupled-cluster expansion, and only an exact treatment of
the strong correlation recovers the true profile.
The same physics, compressed to four electrons, defeats
single-reference methods at multireference enzymatic
transition states and motivates the model studied here.

We focus computationally on the methodological breakdown
at a multireference transition state, using a minimal
surrogate system for which the exact energy is available
as a reference.
To isolate this breakdown we study the rectangular-to-square
scan, see Fig.~\ref{fig:h4-geometry}.
Four hydrogen atoms are placed at the corners of a rectangle
parameterized by angle~$\theta$, with each atom at distance
$R = 1.2$~\AA\ from the center at positions
$(\pm R\cos\theta,\;\pm R\sin\theta)$.
At $\theta < 45^\circ$ the shorter vertical separation
favors one H$_2$ pairing; at $\theta > 45^\circ$ the
horizontal pairing dominates.
At the square geometry $\theta = 45^\circ$ both pairings
are degenerate and the ground state is an equal superposition
of two configurations, the hallmark of strong static
correlation.

The square geometry is the transition state for the
H$_2$ + H$_2$ bond-exchange reaction; the resulting
two-configuration wavefunction more broadly models any
enzymatic TS in which bond homolysis contributes significant
biradical character.
The label biradical here does not signal an intractable
classical problem: the static correlation at the H4
saddle point is fully captured by a (2,2) active
space~\cite{Chan2024}, a size that remains entirely
controllable on classical hardware.
For H4 specifically, CASSCF(2,2) is the minimal classical
remedy that captures this two-configuration character,
though such multireference active spaces remain tractable
only for small surrogates~\cite{Chan2024}.

\section{Methods}
\label{sec:methods}

The H4 scan is computed in vacuum with a family of
electronic structure methods spanning the cost-accuracy
curve from exact to practical~\cite{Chan2024}, extended
here to expose methodological breakdown against an exact
FCI reference.
All methods use the STO-3G basis set; more elaborate
basis sets are applicable in principle, but since the
multireference character at the transition state is
symmetry-driven, STO-3G is sufficient to expose the
qualitative breakdown of single-reference methods.
Full configuration interaction (FCI) provides the exact
energy within this basis and serves as the reference
throughout the scan.
Restricted Hartree-Fock (RHF) is the single-determinant
baseline; second-order M\o{}ller-Plesset perturbation
theory (MP2) adds the leading dynamical correlation
correction on top of RHF.
Density functional theory calculations are performed with
two functionals: B3LYP~\cite{Becke1993}, a global hybrid that has become a
standard workhorse for organic chemistry; and
$\omega$B97M-V, a range-separated meta-GGA with non-local
VV10 dispersion correction.
Coupled-cluster theory is applied at the CCSD(T) level.
The proprietary PexMachina solver targets the exact
ground-state energy without the single-reference constraint;
it reproduces the FCI result throughout the scan,
including at the multireference transition-state geometry
This problem size is well within the operational range
of the PexMachina solver; its inclusion here serves the
purpose of validating the approach against an exact
reference.

The methods span many orders of magnitude in computational
cost and memory.
DFT with a semilocal functional scales as
$\mathcal{O}(N^3)$ in time with the number of basis
functions $N$; the hybrid functionals used here formally
raise this to $\mathcal{O}(N^4)$ due to exact-exchange
evaluation, while memory scales as $\mathcal{O}(N^2)$.
CCSD(T) has a formal time complexity of
$\mathcal{O}(N^7)$~\cite{Raghavachari1989} and memory
$\mathcal{O}(N^4)$ (dominated by the doubles amplitudes
$t_{ij}^{ab}$), which limits canonical implementations to
20\textendash30 heavy atoms in practice.
FCI is exact within the basis set but its Hilbert-space
dimension grows combinatorially as
$\binom{N}{n_\uparrow}\binom{N}{n_\downarrow}$, scaling
exponentially with both orbital count and electron number,
which restricts practical calculations to at most
18\textendash22 electrons in a comparable number of
orbitals~\cite{Knowles1984}.
The PexMachina solver represents the wavefunction efficiently requiring
$\mathcal{O}(M^3 N)$ time and $\mathcal{O}(M^2 N)$ memory
~\cite{White1992,Schollwock2011}, where $M$ is a measure for entanglement;
for ground states satisfying an entanglement area
law---as the H4 ground state does away from the square
geometry---$M$ converges rapidly and the cost is polynomial
in system size.

Multireference character is assessed through the FCI
natural-orbital occupations, computed at each geometry.
For a single-determinant ground state the leading
occupation $n_1$ equals 2 and the lowest $n_4$ equals 0;
deviations from these ideal values quantify how much
weight the exact wavefunction places on configurations
beyond the RHF reference.

One caveat is worth flagging.
Modern computational studies embed the substrate
directly in a slice of the protein using QM/MM methods
and simulate the enzymatic environment explicitly, rather
than inferring the TS geometry from kinetic isotope
effects.
Both approaches target the same quantity: the geometry of
the enzyme's transition state that the inhibitor must
mimic.
The calculations in this work are performed in vacuum and
do not include the protein environment; this choice
isolates the difficulties that electronic structure methods
face at the gas-phase level, before enzyme embedding
introduces additional complexity.

\begin{figure}[tbp]
\centering
\includegraphics[width=\columnwidth]{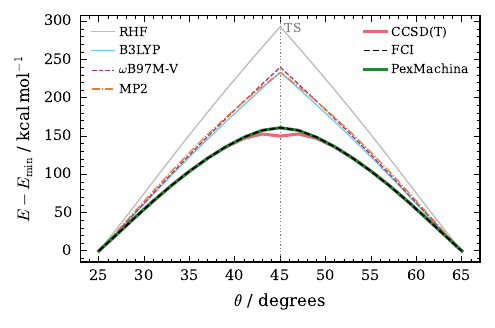}
\caption{%
  Single-reference methods fail catastrophically at the
  H$_4$ transition state.
  Energies are shown relative to the FCI global minimum
  (the equilibrium rectangular geometry) as a function
  of~$\theta$; chemical accuracy is 1~kcal/mol.
  The square geometry ($\theta = 45^\circ$, grey dotted
  line) is the transition state for H$_2$ + H$_2$ bond
  exchange.
  DFT and RHF/MP2 overestimate the barrier and miss
  the chemical accuracy threshold.
  Most critically, CCSD(T) falls \emph{below} the
  exact FCI energy at $\theta = 45^\circ$, returning
  an energy below the exact ground state and marking
  an unambiguous failure of the non-variational
  coupled-cluster expansion.
  PexMachina reproduces the FCI energy at every geometry,
  within chemical accuracy.
}
\label{fig:h4-energy-scan}
\end{figure}

\section{Results}
\label{sec:results}

The H4 scan exposes the breakdown of single-reference methods
at a multireference transition state, with FCI providing the
exact reference throughout.
Figure~\ref{fig:h4-energy-scan} shows the energy of each
method relative to the FCI minimum as $\theta$ is swept
from $25^\circ$ to $65^\circ$.
Away from the square geometry all methods track FCI
reasonably, but near $\theta = 45^\circ$ single-reference
methods diverge sharply.
DFT functionals overestimate the barrier by 42\textendash74~kcal/mol;
RHF and MP2 by $\sim\!178$ and $\sim\!91$~kcal/mol,
respectively.
Most diagnostically, CCSD(T) falls
$\sim\!11$~kcal/mol \emph{below} the exact FCI
energy at the TS, an artifact of the non-variational
coupled-cluster expansion breaking down when the
single-reference state is qualitatively wrong.
PexMachina reproduces the FCI curve exactly throughout.

\begin{figure}[tbp]
\centering
\includegraphics[width=\columnwidth]{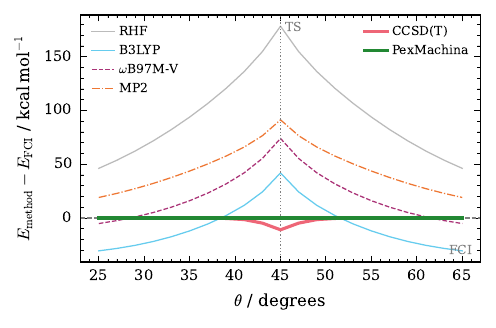}
\caption{%
  At the transition state, the single-reference hierarchy
  inverts: CCSD(T) is further from the exact answer than
  MP2, and there is no internal signal of the failure.
  Signed error $E_\text{method}(\theta) - E_\text{FCI}(\theta)$
  for each method as a function of~$\theta$.
  Away from the square, errors are small but nonzero
  for all methods.
  At $\theta = 45^\circ$, CCSD(T) reaches
  $\sim\!{-11}$~kcal/mol, eleven times the chemical-accuracy
  threshold and negative: the non-variational CCSD expansion
  has overshot and the energy falls below the exact FCI
  ground state.
}

\label{fig:h4-error-fci}
\end{figure}

Figure~\ref{fig:h4-error-fci} isolates the per-geometry
error of each method relative to FCI.
The deviations are small but nonzero at the rectangular
geometries and grow sharply as the square is approached,
confirming that the breakdown is localized to the strongly
correlated TS.
CCSD(T) changes sign from positive at intermediate angles
to negative at the square, illustrating that methods higher
in the single-reference hierarchy can be further from the
exact answer than lower-accuracy ones, with no internal
signal of this failure in advance.

\begin{figure}[tbp]
\centering
\includegraphics[width=\columnwidth]{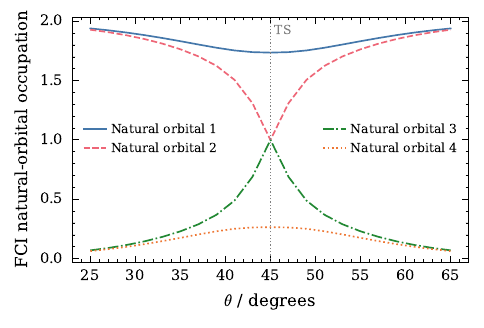}
\caption{%
  FCI natural-orbital occupations diagnose the onset of
  multireference character across the H$_4$ scan.
  The leading occupation $n_1$ and the lowest occupation
  $n_4$ are plotted as a function of~$\theta$; for a
  single-determinant ground state these would be exactly
  2 and 0, respectively.
  Away from the square geometry both occupations remain
  close to these ideal values, confirming that single-reference
  methods are reliable there.
  At the transition-state geometry $\theta = 45^\circ$
  the occupations shift to $n_1 = 1.74$ and $n_4 = 0.26$:
  the exact wavefunction carries 26\% weight on a second
  configuration, so no single Slater determinant is an
  adequate reference, which is the direct cause of the
  CCSD(T) breakdown visible in
  Figs.~\ref{fig:h4-energy-scan} and~\ref{fig:h4-error-fci}.
}
\label{fig:h4-natural-occ}
\end{figure}

Figure~\ref{fig:h4-natural-occ} shows the FCI
natural-orbital occupations as a function of~$\theta$.
At the rectangular geometries $n_1 \approx 2$ and
$n_4 \approx 0$, consistent with a single-determinant
ground state.
At $\theta = 45^\circ$ these shift to $n_1 = 1.74$ and
$n_4 = 0.26$: the exact wavefunction is a near-equal
superposition of two configurations, and no single Slater
determinant can serve as an adequate reference.
The fractional occupation $n_4$ is the microscopic signature
of the single-reference breakdown seen in
Figs.~\ref{fig:h4-energy-scan} and~\ref{fig:h4-error-fci}.
PexMachina natural-orbital occupations are identical,
within numerical accuracy, to the FCI values throughout
the scan, confirming that the
PexMachina solution is exact within this basis.

The H4 scan directly demonstrates the methodological
breakdown at the multireference TS.
A complementary failure mode, cost scaling, afflicts even
weakly correlated systems where CCSD(T) is methodologically
correct but prohibitively expensive~\cite{Chan2024}.
Quantum phase estimation is agnostic to multireference
character and targets the exact ground-state energy directly;
it is among the few approaches that are simultaneously free
of reference-state bias and, in principle, polynomial in
system size.

\section{Conclusion}
\label{sec:discussion}

The H4 rectangular-to-square scan is the most elementary
case of the methodological breakdown studied here: four
electrons in an STO-3G basis, an analytically tractable
active space, and an exact FCI reference at every geometry.
That the proprietary PexMachina solver reproduces this
reference demonstrates that strong correlation at a
multireference TS can be treated exactly, a capability
that single-reference methods fundamentally lack.

The PexMachina solver reproduces the exact ground-state
energy at a strongly correlated transition state,
establishing a quantitative baseline; in future work we
plan to apply this approach to real enzymatic transition
states.

The precise calculation of enzymatic transition states
motivates the use of quantum computation in this context.
Classical methods for strongly correlated systems pay a
high polynomial cost in inverse error: achieving chemical
accuracy in systems that require both strong and weak
correlation simultaneously is exactly the regime where
no classical approach offers a reliable, systematically
improvable solution at the relevant active-space sizes.

As an example, quantum phase estimation targets the exact
ground-state energy with Heisenberg-limited error scaling,
meaning the number of queries grows as $1/\epsilon$ rather
than the $1/\epsilon^2$ of classical Monte Carlo, and is
agnostic to multireference character by construction.
Provable error scaling is the concrete argument for
quantum utility in TS energetics, precisely on problems
where classical methods pay a steep polynomial penalty.
This argument must be evaluated non-asymptotically.
At the system sizes and accuracies relevant to enzymatic
TS computation, prefactors and constant-factor overheads
determine whether any quantum advantage is real.
Progress requires both algorithmic improvements that
reduce these prefactors and experimental demonstrations
on progressively larger active spaces where the crossover
with classical methods can be measured quantitatively.
Future work will extend this approach to further use cases
for quantum computation in drug discovery and lead
optimization.

\bibliography{refs}

\end{document}